
\input harvmac
\noblackbox
\sequentialequations
%

%
%

\def\Title#1#2{\ifx\answ\bigans \nopagenumbers
\abstractfont\hsize=\hstitle\rightline{#1}%
\vskip .5in\centerline{\titlefont #2}\abstractfont\vskip .5in\pageno=0
\else \rightline{#1}
\vskip .8in\centerline{\titlefont #2}
\vskip .5in\pageno=1\fi}
\ifx\answ\bigans

scaled\magstep3
\else

scaled\magstep3
 
 \font\absi=cmmi10 scaled\magstep1
\font\absis=cmmi7 scaled\magstep1 \font\absiss=cmmi5 scaled\magstep1
\font\abssy=cmsy10 scaled\magstep1 \font\abssys=cmsy7 scaled\magstep1
\font\abssyss=cmsy5 scaled\magstep1 
\skewchar\absi='177 \skewchar\absis='177 \skewchar\absiss='177
\skewchar\abssy='60 \skewchar\abssys='60 \skewchar\abssyss='60
\fi
%
%

\def\ajou#1&#2(#3){\ \sl#1\bf#2\rm(19#3)}

\def\frac#1#2{{#1 \over #2}}

\def\ls{{\lambda\sigma}}

\def\vx{{\vec x}}

\def\R{\hbox{\rm I \kern-5pt R}}
\def\ajou#1&#2(#3){\ \sl#1\bf#2\rm(19#3)}

\def\l3{\lambda_3}

\hyphenation{par-am-et-rised}

\def\kbh{\kappa _{bh}}
\def\kds{\kappa _{dS}}
\def\bwwp{\beta _{\omega , \omega '}}
\def\awwp{\alpha _{\omega ,\omega '}}
\def\bwwpbh{\beta ^{bh}_{\omega ,\omega '}}
\def\bwwpds{\beta ^{dS}_{\omega ,\omega '}}
\def\nw{N_{\omega}}
\def\rbh{r_{bh}}
\def\rds{r_{dS}}
\def\uds{U_{dS}}
\def\ubh{U_{bh}}
\def\pw{\phi _{\omega}}
\def\ghm{\hat{\nabla}_\mu}
\def\epb{\bar{\epsilon}}
\def\ep{\epsilon}
\def\epo{\epsilon _o}
\def\gm{\nabla _\mu}

\def\gamth{\gamma^{\hat{t}}}
\def\gam5h{\gamma^{\hat{5}}}
\def\vx{\vec{x}}
\def\muh{\hat{\mu}}
\def\nuh{\hat{\nu}}
\def\hatt{\hat{t}}
\def\hk{\hat{k}}

%
%
\lref\kttodo{D. Kastor and J. Traschen, work in progress.}
\lref\gh{G.W. Gibbons and S.W. Hawking, {\it Phys. Rev.} {\bf D15}, 2738
(1977).}
\lref\romans{L. Romans, {\it Nucl. Phys.} {\bf B393}, 395 (1992).}
\lref\kt{D. Kastor and J. Traschen, {\it Phys. Rev.} {\bf D47}, 5370 (1993).}
\lref\mellor{F. Mellor and I. Moss, {\it Class. Quant. Grav.} {\bf 6},
1379 (1989).}
\lref\ghull{G.W. Gibbons and C.M. Hull, {\it Phys. Lett.} {\bf 109B}, 190
(1982).}
\lref\ghhp{G.W. Gibbons. S.W. Hawking, G. Horowitz and M. Perry, {\it Comm.
Math. Phys.} {\bf 88}, 296 (1983).}
\lref\mpref{S.D. Majumdar, {\it Phys. Rev.} {\bf 72}, 930 (1947); A.
Papapetrou,
{\it Proc. Roy. Irish Acad.} {\bf A51}, 191 (1947).}
\lref\kallosh{R. Kallosh, {\it Phys. Lett.} {\bf 282B}, 80 (1992).}
\lref\parker{L. Parker, in {\it Asymptotic Structure of Spacetime}, New York
(1977).}
\lref\abdes{L. Abbott and S .Deser, {\it Nuclear Physics} {\bf B195}, 76
(1982).}
\lref\jt{J. Traschen, {\it Phys. Rev.} {\bf D31}, 283 (1985).}
\lref\bd{N.D. Birrell and P.C.W. Davies, {\it Quantum Fields in Curved
Spactime}, Cambridge University Press (1982).}
\lref\wald{R.M. Wald, {\it General Relativity}, University of Chicago Press
(1984).}
\lref\ghhp{G.W. Gibbons, S.W. Hawking, G. Horowitz and M. Perry,
{\it Comm. Math. Phys.} {\bf 88}, 295 (1983).}
\lref\ls{D. Kastor and J. Traschen, {\it Phys. Rev.} {\bf D47}, 480 (1993).}
\lref\mcv{G.C. McVittie, {\it Mon. Not. R. Astron. Soc.} {\bf 93}, 325 (1933).}
\lref\witten{E. Witten, {\it Comm. Math. Phys.} {\bf 80}, 381 (1981).}
\lref\wilczek{C. Holzhey and F. Wilczek, {\it Nucl. Phys.} {\bf B380}, 447
(1992).}
%
%
%
%
\Title{\vbox{\baselineskip12pt
\hbox{UMHEP-399}
\hbox{gr-qc/9311025}
}}
{\vbox{\centerline{Particle Production and Positive Energy Theorems for}
\vskip2pt\centerline{Charged Black Holes in DeSitter }}}
\baselineskip=12pt
\centerline{David Kastor\foot{\it Internet: kastor@phast.umass.edu}
and Jennie Traschen\foot{\it Internet: lboo@phast.umass.edu}}
\bigskip
\centerline{\sl Department of Physics and Astronomy}
\centerline{\sl University of Massachusetts}
\centerline{\sl Amherst, MA 01003-4525}
\bigskip
\centerline{\bf Abstract}
We study quantum mechanical and classical stability properties of
Reissner-Nordstrom deSitter (RNdS) spacetimes, which describe black holes
with mass $M$ and charge $Q$ in a background
with cosmological constant $\Lambda \ge 0$.
There are two sources of particle production in these spacetimes;
the black hole horizon
and the cosmological horizon.  A scattering
calculation is done to compute the Hawking radiation in these spacetimes.
We find that the flux from the black hole horizon equals the flux
from the cosmological horizon, if and only if $|Q|=M$, indicating that
this is a state
of thermodynamic equilibrium. The spectrum, however, is not thermal.
We also show that spacetimes containing a
number of charge equal to mass black
holes with $\Lambda \ge 0$, have supercovariantly constant spinors,
suggesting that they may be
minimum energy states in a positive energy construction.
As a first step in this direction, we
present a positive energy construction for asymptotically
deSitter spacetimes with vanishing charge.  Because the construction
depends only on a spatial slice, our result also holds for spacetimes which
are asymptotically Robertson-Walker.


\Date{11/93}


\newsec{Introduction}

In this paper we will discuss quantum mechanical and classical
aspects of the stability of Reisner-Nordstrom deSitter (RNdS) spacetimes,
in which the magnitude of the electric
charge $Q$ is equal to the mass $M$.  In the remainder of the
introduction, we give the argument, based on euclidean quantum field theory,
that $|Q|=M$ RNdS black holes are stable end points of the process of Hawking
evaporation.  In section 2, we present a lorentzian scattering calculation
of particle production in RNdS spacetimes, which shows that the flux of
particles
across the cosmological horizon is equal to the flux across the black hole
horizon, if and only if $Q=M$.  Such a calculation is useful for a number of
reasons.
As well as confirming the Euclidean argument, in this case, it provides
information on
the spectrum of particles produced.  Moreover, generically a spacetime will not
have a real euclidean section, and we will be limited to lorentzian techniques.
 The
multi-black hole generalizations of the $|Q|=M$ RNdS spacetimes \kt\ are
examples of
nonstatic spacetimes which should have interesting thermodynamic
properties\foot{The euclidean picture of black hole thermodynamics can also be
misleading,
as in the case of charged dilaton black holes \wilczek }.
In section 3, we show that $|Q|=M$ RNdS spacetimes admit
spinor fields which are constant with respect to a certain super-covariant
derivative
operator.  The choice of derivative operator is motivated by supergravity
and should lead to a version of the positive energy theorem relevant to RNdS
black holes \kttodo.  Here, we present a more limited, preliminary
result relevant to positivity of a suitably defined energy in
asymptotically Robertson-Walker spacetimes without charge.\foot{ It has
come to our attention that work on a positive  energy construction in
asymptotically deSitter spacetimes has recently been done by T. Shiromizu.}

RNdS spacetimes have metric and gauge field given in static coordinates by
\eqn\rnds{
\eqalign{
ds^2=- f(R)  dT^2 &
+ f(R)^{-1} dR^2 +R^2 d\Omega^2,
\qquad A=-{Q\over R}dT,\cr
&f(R)=1-{2M\over R} +{Q^2\over R^2} -{1\over 3}\Lambda R^2\cr}
}
where the cosmological constant $\Lambda$ is assumed to be positive.
For a range of values of $Q$ and $M$, the spacetime has three Killing horizons;
inner and outer black hole horizons and a cosmological horizon.  The extremal
limit, in which the inner and outer black hole horizons become coincident,
occurs for $M\le |Q|$, with equality in the case $\Lambda =0$.

There are two sources of particle production in
a RNdS spacetime; the black hole horizon and
the deSitter horizon  \gh .
It is interesting to ask whether these two sources can ever be
in a state of thermal equilibrium.  The deSitter horizon would then be
like the wall of a box containing a black hole in thermal equilibrium with a
bath of radiation.
Several facts
suggest that such an equilibrium exists when the magnitude of the
charge $|Q|$ and mass $M$ of the black hole are equal  \mellor\  and
furthermore
that the equilibrium is stable, as we now review.

One can assign temperatures $T_{bh}$ and $T_{ds}$
to the black hole and deSitter horizons
based on the periodicity  that must be imposed on the imaginary time coordinate
in order to obtain a regular euclidean section of the metric \rnds\
at the given horizon  \gh .
The temperatures defined in this way are given by
\eqn\surface{
2\pi k T_{bh}=\kbh,\qquad 2\pi k T_{ds}=\kds,
}
where $k$ is Boltzmann's constant and
$\kbh$ and $\kds$ are the magnitudes of the
surface gravities at the black hole and
deSitter horizons respectively.
For $|Q|=M$, one finds $\kbh  =\kds $ and the two temperatures are equal.
Further, if $M>|Q|$ then one finds that
$\kbh >\kds$ (and vice versa), suggesting that the black
hole radiates (or absorbs) energy to reach equilibrium at $|Q|=M$.
This is also the configuration of maximal gravitational entropy.
The gravitational entropy is given in terms of the areas $A_{bh}$ and $A_{ds}$
of the black hole and deSitter horizons by
$S={1\over 4}\left( A_{bh}+A_{dS}\right) $.
For an infinitesimal perturbation between RNdS solutions with fixed charge
the first law of thermodynamics \gh  states that
$-\kappa _{dS}\delta A_{dS} =\kappa _{bh}\delta A_{bh}$.
\foot{This can be
generalized to include arbitrary, not necessarily
stationary, stress energy perturbations, by the
methods of  \jt . Then there appears an additional volume integral
of the perturbed energy density on the right hand side.}  Hence
$\delta A_{tot}=\delta A_{bh} (1-{{\kbh}\over{\kds}}) =0$
for $\kbh =\kds$.

These arguments based on euclidean quantum field theory
suggest that if quantum mechanical processes are taken
into account, so that for example the area of the black hole horizon can
decrease, then a charged black hole with $|Q|\neq M$ will evolve to
a state with $|Q|=M$. In this paper,
we give the results of a Lorentzian scattering calculation
of the particle production
in Hawking radiation. We compare the flux of particles crossing the
cosmological horizon at late times ( {\it i.e.}, particles
``coming from the black hole'')
to the flux which crosses the black hole horizon ({\it i.e.}, particles
``coming from the deSitter horizon''). We find that the fluxes are equal
if and only if $|Q|=M$. In this sense, a $|Q|=M$ black hole is in equilibrium
with the deSitter background, in agreement with the Euclidean picture.

\newsec{Particle Production}

We compute the rate of particle production in the RNdS spacetimes
(1), for a massless scalar field $\phi$,
by finding the mode mixing coefficients $\bwwpbh$ and $\bwwpds$
 \parker . The conformal diagram for the
relevant portion of RNdS is shown in
\fig\only{Shown is the conformal diagram for the section of
RNdS relevant to the particle production calculation. The region is bounded
by the white hole, black hole, past deSitter, and future deSitter horizons.
Also indicated are the choices for positive frequency waves at each horizon.
To compute $\bwwpbh$, one studies wave propagation along a null geodesic
$\uds=constant$, parallel to the blackhole horizon. The wave is taken to
be positive frequency and outgoing as it crosses the future deSitter horizon.
Propagating back to the white hole horizon, the wave is decomposed into
positive
and negative frequency parts with respect to the coordinate $\ubh$ on the
white hole horizon. Near the white hole horizon the wave vanishes for $\ubh
>0$,
since that is inside the black hole. Similiary, to compute $\bwwpds$,
one considers a wave propagating on a null geodesic $V_{bh} =constant$,
parallel
to the future deSitter horizon, for a wave which is positive frequency and
ingoing at the black hole horizon.}.
The region is bounded
by the white hole, black hole, past and future deSitter horizons,
which have replaced past and future null infinity in the case
with zero cosmological constant.
The first step is to define what is meant by particles, {\it i.e.} we must
choose a time coordinate near each boundary, which defines the
positive frequency modes there. Define the following Kruskal
type coordinates which are well behaved near the appropriate
horizons
\eqn\kruskal{\eqalign{
\ubh =-{1\over {\kbh}}e^{-\kbh u}   ,&\qquad V_{bh} ={1\over {\kbh}}
e^{\kbh v}\cr
\uds={1\over \kds}e^{\kds u} ,&\qquad V_{dS} =-{1\over \kds}e^{-\kds v}  }
}
where, in standard notation $u$ and $v$ are defined in terms of the tortoise
coordinate $r^*$ according to
\eqn\tortoise{
u=t-r^* \ , v=t+r^* \  , dr^* ={{dr}\over {f ^2}} .
}
Near the black hole horizon, the metric then has the limitting form
\eqn\near{
ds^2 \approx \kbh d\ubh dV_{bh},
}
showing that these are good cooordinates near the black hole horizon.
One can similarly show that the coordinates are also good near
the deSitter horizon.

For $\Lambda =0$, one defines positive
frequency near $I^-$ by $\phi \sim e^{-i\omega v}$,
and positive frequency on the white hole horizon by
$\phi \sim e^{-i\omega \ubh}$.
This choice of positive frequency
on the white hole horizon correctly gives the outgoing thermal
flux at $I^+$,  allowing one to replace the collapsing body in Hawking's
original calculation with this boundary condition on
the field (see e.g.  \wald ,\bd ). This suggests that for
RNdS spacetimes, positive frequency at each horizon should be defined with
respect to
the appropriate Kruskal coordinate there. One can check that
this is equivalent to choosing the affine parameter on {\it e.g.}, the
white hole horizon to define positive frequency there.

The Klein-Gordon equation for $\phi$ near any of the horizons reduces to
the free wave equation.  As in the case with no cosmological constant,
the potential term due to the background gravitational field decays
exponentially in $r^*$ near a horizon.  Consider then
a pure positive frequency outgoing wave near the deSitter horizon
at late conformal time, $\pw\sim e^{-i\omega \uds}$
(see \only ). In the geometrics optics
limit (as with $\Lambda =0$), finding the form of this wave
propagated back to the white hole horizon reduces to finding the
dependence of the coordinate $\uds$ on the coordinate $\ubh$.
Using \kruskal , it then follows that on the white hole horizon
\eqn\whitehole{
G_{\omega}(\ubh )\equiv
\pw (\uds (\ubh ))\sim \left\{
\eqalign{
e^{-i\omega \xi ^2 ({-1\over \ubh})^{\eta}} ,\qquad &
\ubh <0\cr    0, \qquad & \ubh >0\cr} \right.
}
where $\eta \equiv \kds /\kbh$ and $\xi ^2 \equiv {1\over \kds }
({1\over \kbh })^{\eta} $.
The particle production follows from determining the negative frequency
portion of this wave on the white hole horizon,
\eqn\betabh{
\bwwpbh  =\left |{{\omega '}\over {\omega}}\right |^{{1\over 2}}
\int d\ubh e^{-i\omega^\prime \ubh} G_{\omega} (\ubh )
}

Similiarly, there is emission ``from'' the deSitter horizon as seen
by an observer outside the black hole horizon at late times.  Consider
a positive frequency wave which is entering the black hole horizon (see
\only ), $\pw \sim e^{-i\omega V_{bh}}$. Then in the geometrics optics
approximation, on the past deSitter horizon, the wave is given by
\eqn\desitter{
F_{\omega}(V_{dS}) \equiv \pw (V_{bh} (V_{dS}) \sim \left\{
\eqalign{
e^{-i\omega\mu ^2 ({-1\over V_{dS}})^{{1\over \eta}}},\qquad & V_{dS} < 0\cr
 0, \qquad & V_{dS} >0\cr} \right.
}
where $\mu ^2 =1/\kbh (1/\kds )^{{1\over \eta}}$.
Similarly to \betabh , $\bwwpds$ is given in terms of the fourier transform
of \desitter .
For general values of $Q$ and $M$, the functions $F_\omega$ and $G_\omega$
appearing in \whitehole\ and \desitter\ are related according to
\eqn\relation{
G_{\omega}(x) =F_{{\omega\over\eta ^2}}
(x^{\eta ^2}).
}
We see that the two functions are equal for
$\eta =1$, which occurs when $|Q|=M$.  We then have the main result of this
section
\eqn\equality{
\bwwpbh =\bwwpds \qquad {\rm if\ and\ only\ if} \quad |Q|=M.
}
This implied that for each horizon, the flux of particles absorbed is equal to
the
flux of particles absorbed.

The spectrum of emitted particles is given by $\nw =\int d\omega '
|\bwwp |^2 $. We can estimate the above integrals using the stationary
phase approximation. It is simpler to work with the coefficients
$\awwp =-\beta _{\omega ,-\omega '}$. The stationary phase approximation
can be used to evaluate this integral, and $\bwwp$ is then gotten
by analytically continuation, after noticing that \whitehole\ implies that
$\awwp$
is analytic in the lower half $\omega '$ plane.

For the case $|Q|=M$, when the surface gravities are equal, one finds in the
stationary
phase approximation
\eqn\cando{
|\bwwp |^2 ={\pi\over {2\kappa}} {1\over{|\omega\omega '|^{1/2}}}
e^{-{4\over{\kappa}}|\omega\omega '|^{1/2}}
}
This approximation can be checked, because in the $|Q|=M$ case
the integral
can be done exactly. One can show that $\awwp$ is only a
function of the combination
$x\equiv {1\over\kappa}|\omega\omega '|^{1/2}$. The function $\alpha (x)$
can then be shown to satisfy the differential equation
\eqn\alphadiff{
\alpha ''(x) +{1\over x}\alpha ' +(4-{1\over x^2})\alpha =0.
}
Hence $\alpha$ is proportional to the Bessel function
$J_1 (x)$. Again analytically continuing, we then have
$\bwwp \propto K_1 ({2\over\kappa}|\omega\omega '|^{1/2}) $.
Using the asymptotic
behavior of the modified Bessel function $K_1(x)$, this agrees with
our previous expression for large values of the argument $x$,
and fixes the constant of
proportionality to be $\sqrt{2}/\kappa$. For small values of the argument,
one then finds
$|\bwwp |^2\approx {1\over 2\omega\omega '}$. However, the
geomtrics optics approximation which was used in doing
the calculation means that the results are only valid
at high frequencies. We would expect that the wavelength of the
propagating wave must be small compared to the length scales $\rbh$ and
$\rds$. Using a lower limit of $2\pi /\rbh$ in the integration over
frequencies, one finds for the spectrum
\eqn\spectrum{
N_{\omega}\approx {\pi\over 4\omega} e^{-4\sqrt{2\pi\omega}\over
\kappa \sqrt{a}}
}
This is not a thermal spectrum, so the situation is probably best
described as a thermodynmic equilibrium. If the lower cutoff were
taken to be proportional to $\omega$, rather than $1/\rbh$, the spectrum
would be thermal.

There are several limits one can take in order to check this result. Letting
$\kappa \rightarrow 0$ above, corresponds to keeping
$|Q|=M$ and letting the cosmological constant
$\Lambda$ approach zero, so that the spacetime approaches
extremal Reissner-Nordstrom. In this limit $N_{\omega}$ goes to zero,
as it should. Secondly, one can consider setting $Q=0$, and then
letting $\Lambda\rightarrow 0$, so that the metric approaches Schwarzschild.
This is equivalent to letting $\rds$ approach zero, while keeping $\rbh$
finite. The particle production \betabh\ from the black hole can again
be evaluated in the stationary phase approximation. One finds that
the coefficients $\awwp$ approach those for a Schwarzschild black hole.

\newsec{Supercovariantly Constant Spinors for $\Lambda >0$}

We now turn to issues relevant to the
classical stability of $|Q|=M$ RNdS spacetimes.
First, recall that for $\Lambda =0$
a generalization of the positive
energy theorem  \ghull,\ghhp\
implies that the ADM mass $M$ of an asymptotically flat spacetime
satisfies the inequality
\eqn\bound{
M\ge |Q|,
}
where $Q$ is the total electric charge of the spacetime.
This inequality is saturated for
spacetimes which admit a spinor field $\epsilon$
satisfying $\hat{\nabla}_\mu\epsilon =0$, where $\hat{\nabla}_\mu$ is a certain
supercovariant derivative operator
arising naturally in ungauged $N=2$ supergravity.  In the context of
supergravity
$\hat{\nabla}_\mu\epsilon$ gives the variation of the gravitino field under an
infinitesimal local supersymmetry transformation.  Spacetimes having
supercovariantly constant spinors have unbroken supersymmetries.
Amongst asymptotically flat
electrovac solutions, the bound \bound\ is saturated by the
Majumdar-Papapetrou multi-black hole solutions  \mpref ,
which represent collections of
extremal ($|Q|=M$) charged black holes.  These spacetimes each admit two
supercovariantly constant spinors.
Calculations of the effective potential in supergravity have shown that
these spacetimes remain ground states at the quantum level as well
\kallosh .

A nonzero cosmological constant $\Lambda=-3g^2$
arises in {\it gauged}
$N=2$ supergravity, where $g$ is the coupling constant of the
gravitino with the $U(1)$ gauge field.  If the coupling $g$ is real, as one
would require in a quantum theory, then the cosmological constant is negative.
Romans  \romans\  has written down
the supercovariant derivative operator $\hat{\nabla}_\mu$,
which arises in gauged $N=2$
supergravity, with real coupling $g$ and showed that $|Q|=M$ RNadS spacetimes
each admit two
supercovariantly constant spinors\foot{These spacetimes have naked
singularities, because for $\Lambda<0$ the extremal limit occurs with
$|Q|<M$.}.
At the classical level one can also consider imaginary
values of $g$ and hence positive values of the cosmological constant.
With the substitution $g=iH$, the supercovariant derivative operator is given
by
\eqn\supercov{
\hat{\nabla}_\mu\epsilon  =  \left( {}^4\nabla_\mu +HA_\mu +
{i\over 2}H\gamma_\mu  +
{i\over 4}F_{\rho\sigma}\gamma^\rho\gamma^\sigma\gamma_\mu \right)\epsilon
}
We are then interested whether or not
$|Q|=M$ RNdS black holes admit spinor fields
satisfying $\hat{\nabla}_\mu\epsilon =0$.
If this is the case, then it seems promising
that a version of the positive energy theorem can be proved, showing the
classical
stability of these spacetimes.

In fact we can find supercovariantly constant spinors for a more general class
of
spacetimes, which includes the $|Q|=M$ RNdS spacetimes.  These are the
deSitter analogues of the MP multi-black hole solutions, which were found in
\kt.
In spatially flat Robertson-Walker type coordinates, these spacetimes
are given by
\eqn\ktmetric{ \eqalign{
ds^2 = -{1\over\Omega(\vx)^2}dt^2 + a(t)^2\Omega(\vx)^2 &
\delta_{ij}dx^i dx^j ,\qquad
A={1\over \Omega}dt,\cr
\Omega = 1+\sum_i {M_i\over a|\vx -\vx_i|},&\qquad a(t)=exp(Ht).\cr  }
}
We find that these spacetimes each have two supercovariantly constant
spinors, which are given by
\eqn\spinor{
\epsilon = {1\over \sqrt{a\Omega}}\epb = {e^{-Ht/2}\over\sqrt{\Omega}}\epb
}
where $\epb$ is a constant spinor satisfying the
projection\foot{In our conventions hatted indices denote frame indices
and unhatted indices are coordinate indices.  Hence
$\{\gamma^{\muh},\gamma^{\nuh}\}=2\eta^{\muh\nuh}$ and
$\{\gamma^{\mu},\gamma^{\nu}\}=2g^{\mu\nu}$.}.
$(1-i\gamth) \epb = 0$

\newsec{Positive Energy Constructions}

As stated above, the supercovariant derivative operator \supercov\ should
be the starting point for a positive energy construction, showing that a
suitably defined mass for asymptotically deSitter spacetimes is minimized for
fixed
charge by the spacetimes \ktmetric .
The relevant
mass in the asymptotically deSitter case should turn out to be that given by
Abbott and Deser in \abdes .
In the present note,
we will address only the case $Q=0$.
Such a construction for $\Lambda<0$ is given in
\ghhp .
We will return to the $Q\neq 0$ case in a
future publication \kttodo .
The result we present here, however, will be valid for spacetimes
which are asymptotic to spatially flat Robertson-Walker,
not necessarily asymptotically deSitter. This is because the construction
takes place only on a spatial slice, and a spatial slice of Robertson-Walker
can be
embedded in a deSitter spacetime.
In this case the asymptotic mass will be related to the
boundary  term given in \jt , in the context of peturbations of
Robertson-Walker spacetimes,
and discussed further in \ls .

Let the metric $g_{\mu\nu}$ of a spacetime asymptotically
approach the metric for a Robertson-Walker spacetime
$g_{\mu\nu}^{RW}$ on each $t=constant$ spatial slice, where
\eqn\rw{
ds^2_{RW}=g_{\mu\nu}^{RW}dx^\mu dx^\nu
=-dt^2 +a^2 (t)\delta _{ij} dx^i dx^j .
}
With vanishing gauge field the supercovariant
derivative operator \supercov\ becomes %
\eqn\rwspinor
{\ghm \ep =\gm \ep +{i\over 2} H(t) \gamma _a \ep,
}
where $H(t)\equiv \dot{a}(t) / a(t)$ may depend on time.

When the metric is exactly Robertson-Walker, there exist four supercovariantly
constant
spinors $\ep _o$ satisfying $\hat{\nabla}_\mu
\ep _o =0$.
This can be seen as
follows. In five dimensional Minkowski spacetime there are four
constant spinor solutions to $\nabla _\mu \psi =0$. Projecting this
equation onto the four dimensional deSitter hyperboloid, one finds
that there are four solutions to the equation $\tilde{\nabla }_{\mu} \psi
\equiv
(\gm +{H\over 2}\gamma _a \gam5h ) \psi =0$. Letting $\epo =(1-i\gam5h )
\psi $, we have\foot{Note that $(\gamma^{\hat{5}})^2=+1$,
so this is not a projection.}
$\ghm \epo =0$. A $t=constant$ spatial
slice in a general Robertson-Walker spacetime \rw\ can be embedded in the
deSitter hyperboloid with the appropriate value of $H(t)$.
On a given spatial slice then, there are four
solutions to $\hat{\nabla} _j \epo =0$ , where $j$ is a spatial index.
Explicity, these are given by
\eqn\newspinor
{\epo =(1-i\gam5h )(1-{H\over 2} x^i \gamma _i (\gamth +\gam5h ))\epb
}
where $\epb$ is any constant spinor.

Since the positive energy construction will take place on a spatial surface,
this is actually sufficient to construct a positive energy statement,
for any spatial slice in the coordinates of equation \rw .
However,
there are solutions which satsify the full equation
$\hat{\nabla}_{\mu} E =0$.
These are conveniently given by $E =exp(\pm \int ^t
{H\over 2}dt' )\epo $, where $\gamth \epo =\mp i \epo$. The latter
projection is equivalent to $(\gamth \pm \gam5h )\epb =0$.

Let $g_{ij}$ and $K_{ij}$ be the metric and extrinsic curvature
on a spatial slice, which approach the metric and extrinsic curvature
of a Robertson-Walker spatial slice with Hubble constant $H$.
Assume that $g_{ij}$ and $K_{ij}$ solve the Einstein constraint
equations with sources $T_{\hatt\hatt}$ and $T_{\hatt\hk}$.
Let $\ep$ be a solution to the contracted equation
\eqn \dirac {\gamma ^j \hat{\nabla}_j \ep =0 }
on this spatial slice ( where
the index $j$ runs over spatial values only) with the boundary
condition that $\ep$ approaches one of the constant spinors
$\epo$ at large $r$. Let $V$
be a volume in the spatial slice. Then following the analysis of \witten\
we find the following relation
\eqn\positive{
\int _{\partial V}\  da_i \ep ^{\dagger} \hat{\nabla }^i \ep
=\int _V d^3 x\ \sqrt{g}\left (
4\pi\ep ^{\dagger} (T_{\hatt\hatt} -T_{\hatt\hatt}^{RW} + T^{\hat{t}}_k \gamma
^k
\gamth ) \ep \  +(\hat{\nabla}_i \ep )^{\dagger} \hat{\nabla }^i \ep
\right ) ,
}
where $T_{\hatt\hatt}^{RW} =3H(t)^2/8\pi$ is the energy density of a
Robertson-Walker
spatial slice.
If black holes are present in $V$, then the boundary $\partial V$ has
components
at the black hole horizons. These boundary terms
can be made to vanish by imposing a suitable projection
on $\ep$, as shown in \ghhp . This leaves the outer boundary of $V$,
which we will take to be spatial infinity.  In the case $\Lambda=0$, this would
be
related to the ADM $4$-momentum \witten . The volume integral on the right hand
side
of \positive\ can, in the context of asymptotically Robertson-Walker
spacetimes, have
both positive and negative contributions.
The second term, involving derivatives of the spinor field, is positive
definite.  However, the first term, which is a difference between the matter
stress
energy and the background Robertson-Walker energy density, can be either
positive
or negative,
depending on whether there are mass overdensities or underdensities.
If the matter stress energy satisfies the dominant energy condition,
however, this term cannot be
more negative than minus the energy density of the background itself.
Furthermore, for asymptotically deSitter spacetimes it is natural to consider
this
difference to be positive.

One expects that
the boundary term at spatial infinity gives an
analogue of the ADM $4$-momentum for asymptotically Robertson-Walker
spacetimes.
Such a quantity has been defined for asymptotically deSitter spacetimes by
Abbott
and Deser \abdes .
For each asymptotic killing vector in an asymptotically deSitter spacetime, one
can
construct a conserved charge given by a
two dimensional boundary integral on a spatial slice.  These charges all vanish
for
pure deSitter.
Hence, as is argued in the
anti-deSitter case \ghhp , we expect that for asymptotically deSitter
spacetimes, the boundary term in \positive\ gives
the Abbott-Deser mass.
In the example of Schwarzchild-deSitter, one can
evaluate the boundary term explicity, and it gives the mass parameter.

More generally,
the McVittie metric \mcv\ describes a mass embedded in a Robertson-Walker
spacetime,
\eqn\mcvit{
ds^2 =({1- {m\over 2ar} \over 1+{m\over 2ar} })^2 dt^2
+ a^2 (t) (1+{m\over 2ar})^4 \delta _{ij} dx^i dx^j  .
}
In this case, the exact solution to the Dirac equation \dirac\ can be found.
The boundary term can then be evaluated and shown to give
the mass parameter $m$.
Hence, if a metric approaches \mcvit\ at large radius up to terms
of order $1/r^2$, the boundary term will continue to give the appropriate mass
parameter.  In principle, one can find a general expression for the boundary
term
in \positive\
by solving \dirac\ asymptotically on a spatial surface
to sufficiently
high order, as in \witten .  In practice, here, this
is complicated by the nonzero background extrinsic curvature and we have not
yet found
a general expression.
We expect that the final result will be related to the boundary
integral which arises in the analysis of \jt\ on perturbations of
Robertson-Walker
spacetimes.

\bigbreak\bigskip\bigskip\centerline{{\bf Acknowledgements}}\nobreak
JT is supported in part by NSF grant NSF-THY-8714-684-A01.

\listfigs
\listrefs
\end